# It's a Matter of Principle.
# Scientific Explanation in Information-Theoretic Reconstructions of Quantum Theory

*Laura Felline*
*Department of Philosophy, Communication and Media Studies*
*University of Rome Three*

**Abstract**

The aim of this paper is to explore the ways in which Axiomatic Reconstructions of Quantum Theory in terms of Information-Theoretic principles (ARQITs) can contribute to explaining and understanding quantum phenomena, as well as to study their explanatory limitations. This is achieved in part by offering an account of the kind of explanation that axiomatic reconstructions of quantum theory provide, and re-evaluating the epistemic status of the program in light of this explanation. As illustrative cases studies, I take Clifton's, Bub's and Halvorson's characterization theorem and Popescu's and Rohrlich's toy models, and their explanatory contribution with respect to quantum non-locality. On the one hand, I argue that ARQITs can aspire to provide genuine explanations of (some aspects of) quantum non-locality. On the other hand, I argue that such explanations cannot rule out a mechanical quantum theory.

**Introduction**

In the philosophy of science literature, causal explanations have traditionally been of high import, and in Quantum Theory (QT) such explanations require contending with ontological questions. Given the well-known issues in the ontological interpretation of QT, such a requirement has often been considered (by philosophers) an obstacle for the explanation of quantum phenomena. In particular, such a requirement seems to be hardly compatible with the expectation of genuine explanations produced by research programs that remain agnostic about to the ontology of QT. One of the most manifest examples of this difficulty comes from the program of axiomatic reconstruction of QT. Within the current mainstream research in the foundations of QT much attention has been turned to this program, and in particular, to those reconstructions that focus on information-theoretic principles (Clifton et al. 2003, Popescu and Rohrlich 1994, Popescu 2006, Zeilinger 1999, Rovelli 1996). It is therefore becoming increasingly important for philosophers of science to deal with the explanatory gain that is implied in the switch from the traditional interpretation program to the program of Axiomatic Reconstruction of QT in terms of Information-Theoretic principles (ARQIT).

The analysis proposed in this paper begins from the assumption that science provides many varieties of explanation other than causal explanation and that some of these alternative kinds of explanation do not require a commitment to any specific ontology. With this background assumption in mind, I will examine some ways in which ARQIT can contribute to explaining and understanding quantum phenomena, as well as describe their explanatory limitations.

I will therefore first offer an outline of an account of scientific explanation within the context of ARQIT. Afterward I will investigate the legitimate role of ARQIT in the foundations of QT. The literature abounds with thorough analyses about the role of quantum information theory and ARQIT in the foundations of QT and the conclusions I will reach (spoiler: while ARQIT can contribute to some traditional issues in the foundations of QT, it fails to address others. Also, ARQIT does not rule out the possibility or necessity of a traditional interpretation of QT to explain such other unsolved issues) are not new in the literature – they converge, for instance, with the analysis put forward in (Timpson 2013) and partially with (Duwell 2007). What I take to be the novelty of the approach I propose, however, is that it is specifically carried out within the domain of the theory of scientific explanation.

As a concrete illustration of how these theories can contribute to the explanation and understanding of the quantum world, I analyse the accounts ARQITs provide of non-locality, defined minimally as quantum entanglement, yielding non-local quantum correlations in the sense of Bell's theorem. Although the proposed analysis will be centred on two specific theories as case studies, a good part of the analysis here proposed is grounded on generic features of ARQITs and their explanations. For the most part, the conclusions I will reach can therefore be extended, *mutatis mutandis*, to ARQIT in general. As will be shown, ARQIT provides novel and genuine explanations of some aspects of non-locality, but it fails to address what is seen by most as *the* problem of non-locality, i.e. the problem of *how* quantum correlations occur. The first case study (§ 2) comes from the Clifton-Bub-Halvorson (CBH) reconstruction of QT (Clifton et al. 2003 Bub 2000, 2004, 2005), which explains why there are non-local entangled states. The second case study (§ 3) comes from partial reconstructions of QT (Popescu and Rohrlich 1994, Brassard et al. 2006, Brunner and Skrzypczyk 2009), whose aim is to find an answer to the question 'why is our world only *this much* non-local, and not more than that?'.

§ 4 is devoted to a pre-theoretic illustration of the kind of epistemic gain achievable with ARQITs' explanations. In this section I will make explicit some central features for which a successful model of explanation in ARQIT must account. To anticipate part of the contents of this

section, I will argue that ARQIT addresses one of the central questions in the foundations of QT: how does the quantum world differ from the classical one? Elaborating on the comparative kind of understanding that ARQIT provide, I will conclude that ARQIT's explanation of (some aspects of) non-locality corresponds to a very concrete and straightforward pre-theoretic sense of explanation, i.e. that you explain a feature or a behaviour *P* of *s* by showing how *P* depends on the essence of *s*. In § 5 I describe the bare bones of the proposed account of explanation in ARQIT, which will take inspiration from Mark Steiner's (1978) account of explanatory proofs. Furthermore, I will anticipate two possible objections to the proposed account and show how such features are in fact non-problematic.

In § 6 I provide two examples of how taking into account the specific explanatory work at play within ARQIT might contribute with a new perspective to topics and debates in the foundations of QT. As a first example of such a contribution I analyse the claim that the explanatory role of information shows that information has a special role in the ontology or interpretation of QT, and argue that such a claim is unwarranted. As a second example, I analyse the claim that ARQIT rules out the possibility of a traditional ontological interpretation of QT (and, for instance, its commitment to the reality of the collapse of the wave-function). In contrast with such claims I elaborate on the results of the previous sections and argue that, although ARQIT has the potential of providing a successful explanation of some specific aspects of quantum non-locality, it cannot replace traditional ontological interpretations of QT. CBH – especially Bub – have promoted the claim that ARQIT makes traditional interpretations of QT explanatorily irrelevant. Since their arguments have significantly informed the debate over such issues, a fair part of the discussion proposed in this section is therefore devoted to them.

**1. ARQIT: scopes and strategies**

The program of axiomatic reconstruction of QT departs significantly from the aims of the interpretative program. Interpretations of QT typically aim at grounding the formalism of QT on hypothetical assumptions about the constituents of the quantum world and the processes they undergo, answering the question: "how could the world be such that it behaves the way QT predicts?". The focus and strategy of axiomatic reconstructions of QT is drastically different. They aim at finding a few general physical principles from which, once translated into mathematical terms, one can formally derive the structure of QT:

"[t]heorems and major results of physical theory are formally derived from simpler mathematical assumptions. These assumptions or axioms, in turn, appear as a representation in the formal language of a set of physical principles." (Grinbaum 2007, p. 389)

Principles "must be simple *physical statements*, that is assertions, such that their meaning is immediately, easily accessible to a scientist's understanding." (*ibid.*, p. 390) In the absence of additional assumptions, the function of principles at the basis of axiomatic reconstructions is uniquely of providing an axiomatic basis for the deduction of the rest of the theory, and "nothing can be generally said about their ontological content or the ontic commitments that arise from these principles", since they "have only a minimal epistemic status of being postulated for the purpose of reconstructing the theory in question." (*ibid.*, p. 391). Within quantum information theory in general, and in ARQIT specifically, information is a physically defined quantity cashed out in terms of the resources required to transmit messages – measured classically by Shannon entropy or in QT by Von Neumann entropy.

## 2. The CBH characterization theorem

The CBH characterization theorem (Clifton et al. 2003) has long been the most discussed (by philosophers) ARQIT, and many discussions about the philosophical significance of ARQIT focus on it. The CBH theorem presupposes a mathematical framework called C*-algebra which, according to the authors, is neutral enough to allow a mathematically abstract characterization of a physical theory that includes, as special cases, all classical mechanical theories of both wave and particle varieties, and all variations on QT, including quantum field theories. Within such a framework, CBH formulate a theorem characterising QT in terms of three principles about impossibilities of information transfer:

*The impossibility of superluminal information transfer between two physical systems by performing measurements on one of them ('no superluminal information transfer')*. This constraint states that merely performing a local (non-selective)[1] operation on a system *S* cannot convey any information to a physically distinct system. This constraint corresponds to no signalling via entanglement ('no signalling') featuring in ordinary quantum mechanics.

*The impossibility of perfectly broadcasting the information contained in an unknown physical state ('no broadcasting')*. Broadcasting is a generalization of the process of cloning – a process that starts with a system in any arbitrary state $|\mapsto\rangle$ and ends up with two systems, both in the state $|\mapsto\rangle$. While cloning applies only to pure states, broadcasting generalizes also to mixed states, and, for

---

[1] Selective measurements operations are obviously not considered here, given that selection changes the ensemble under study and therefore its statistics.

pure states, reduces to cloning. In quantum mechanics, broadcasting is possible for a set of states $\rho_i$ iff they are commuting. (Barnum et Al. 1996)

*The impossibility of unconditionally secure bit commitment ('no bit-commitment')*. The bit-commitment is a cryptographic protocol that guarantees a secure commitment to a particular value by the first party, where that value is hidden from a second party. We can illustrate the protocol with a simplifying analogy borrowed from Timpson (2013): in the first stage of the protocol, the first party, Alice, writes 0 or 1 in a piece of paper, then she locks the piece of paper in a safe and gives it to Bob. Bob cannot read the content of the information given by Alice until, at the second stage of the process, she gives him the key of the safe. The protocol is said to be secure if neither Alice can cheat by changing what she wrote in the paper after giving the safe to Bob, nor can Bob cheat by reading the value written by Alice before she gives the key to him. A quantum version of the protocol was invented by Bennett and Brassard (1984). In this version of the protocol Alice encodes the 0 and 1 values into two mixtures represented by the same density operator. Given that the two mixtures are indistinguishable, Bob must wait for Alice to tell him what procedure she used, in order to 'read' the value to which she committed. However, Alice could use a so-called EPR cheating strategy: she can give to Bob one of an entangled pair of particles, and keep the other. By doing this, after giving the particle to Bob, she can change the value of Bob's particle by steering Bob's particle into the desired mixture via appropriate measurements on her particle. Bob cannot reveal her cheating and this makes the protocol not secure.[2]

An important part of the CBH paper is devoted to the characterization of QT against classical phase space theories. They first show that classical phase space theories correspond to commutative (i.e. abelian) C*-algebras:

"not only does every classical phase space presentation of a physical theory define a C∗-algebra, but, conversely, behind every abstract abelian C∗-algebra lurks in its function representation a good old-fashioned classical phase space theory. All of this justifies treating a theory formulated in C∗-algebraic language as classical just in case its algebra is abelian. It follows that a necessary condition for thinking of a theory as a *quantum* theory is that its C∗-algebra be non-abelian". (Clifton et al. 2003, p.1568)

Later, they commit to showing that: i) "the 'no superluminal information transfer' condition entails that the C∗-algebras *A* and *B*, whose self-adjoint elements represent the observables of [two spacelike separated systems] *A* and *B*, commute with each other," (*ibid.*, p.1570) i.e., every element of *A* commutes pairwise with every element of *B*, or, in other words, distinct systems are kinematically independent; and (ii) "the 'no-broadcasting' condition entails that *A* and *B* separately are non-commutative (non-abelian)." (*ibid.*) i.e. any individual system's algebra of observables must be non-commutative; non-commutativity of individual algebras is the formal representative of the physical phenomenon of interference.

---

[2] Bub stresses that a secure bit commitment is also impossible for classical systems, but as a consequence of the impossibility in practice to guarantee the security of the protocol due to issues of computational complexity. "No principle of classical mechanics precludes Bob from extracting this information." (Bub 2005, p.553)

Finally, CBH show how quantum non-locality follows from these two joint algebraic features:

"if *A* and *B* are nonabelian and mutually commuting (and C∗-independent), it follows immediately that there are nonlocal entangled states on the C∗-algebra *A* ∨ *B* they generate" (*ibid.*).

More specifically, the explanation of non-locality follows three steps:

1) from 'no superluminal information transfer follows that the commutativity of distinct algebras is guaranteed: if the observables of distinct algebras commute, then the 'no-superluminal information transfer' constraint holds. Commutativity of distinct algebras is meant to represent 'no signalling'. A theory violating this principle would display strong non-locality and superluminal signalling;

2) from 'no broadcasting' follows the non-commutativity of individual algebras. Cloning is always allowed by classical theories and if any two states can be (perfectly) broadcast, then the algebra is commutative. A theory violating this principle is therefore a classical theory with commutative individual algebras;

3) if *A* and *B* are two sub non-commutative and mutually commuting algebras, there are nonlocal entangled states on the C∗-algebra *A* ∨ *B* they generate.

This was an outline of the account of non-locality provided by CBH. It has to be noticed, though, that CBH argue that such derivation is not sufficient to guarantee the existence of non-local entangled states and that 'no bit-commitment' is also necessary. The justification CBH provide for the introduction of the 'no bit-commitment' principle, however, remains dubious. In particular, we have seen that in the context of C*-algebras, non-locality already follows from the conjunction of 'no broadcasting' and 'no superluminal information transfer'. CBH argue that in the context of a weaker algebra (Segal algebra, for instance), these two principles might not be sufficient, and that the role of 'no bit-commitment' is exactly to guarantee non-locality even in these cases. However, while the 'no bit-commitment' principle is not necessary in the context of C*-algebras, there is no proof that in the context of Segal algebra from the three principles non-locality follows. It is for this reason that Timpson (2013) concludes that this principle is either redundant or inefficacious.

The choice of reconstructing the CBH explanation of non-local entanglement as following from the first two principles only allows us to illustrate a feature of this explanation that will be very useful in the following sections. A slightly technical digression on this point is therefore needed. Various philosophers have objected to the CBH reconstruction that the choice of a C*-algebra might fail to represent the neutral formal framework that an axiomatic reconstruction requires. C*-algebras, in fact, define states as linear functionals of observables, and this assumption rules out by fiat non-equilibrium deterministic hidden variables theories (Timpson 2013, ch. 8.3.2.2). The assumption of equilibrium is an additional assumption necessary to deterministic hidden variables theories in order to reproduce the predictions of QT (without such an assumption, for instance, a deterministic hidden variable theory might allow superluminal signalling). However, the fact that C*-algebras rule out by fiat the possibility of non-equilibrium leads to the elimination of an interesting class of theories – which, in turn, spoils the resulting characterization of QT. In fact, by narrowing the range of theories against which QT is picked out, the formal background plays a relevant role in selecting QT – a role that should exclusively be played by the information-theoretic principles.[3]

## 3. Partial reconstructions of QT

In recent years partial reconstructions of QT in terms of information-theoretic principles have been experiencing a vast success, and their progress in the inquiry on quantum non-locality is especially promising. The strategy here is quite different with respect to the one used in the CBH reconstruction. It consists in building models of 'fantasy quantum theories' that replicate some aspects of QT (i.e. that maintain some principles in common with QT), while changing some others. The consequences of these modifications are then studied by manipulation of such toy models. This makes it possible to investigate the logical structure of the theory and in particular, by changing one principle of QT and seeing how one gets new characteristics in the models, to highlight which principle is responsible for a given quantum feature, and how.

It is indeed fascinating and instructive to see how this strategy has led to important achievements in the discovery and explanation of surprising features of quantum non-locality. Notice that, as we are about to see, the explanandum chosen in this case study is not non-locality itself, but the fact that our world instantiates a particular amount of non-locality rather than somewhat more or somewhat less. The ground-breaking work in this research is represented by (Popescu and Rohrlich 1994), which starts with the question: "What is the quantum principle?". While "[i]n the conventional approach to quantum mechanics, indeterminism is an axiom and nonlocality is a theorem" Popescu and Rohrlich make "nonlocality an axiom and indeterminism a theorem" (1994, p.379). This strategy is justified by the fact that "quantum nonlocality seems as fundamental as ever" and "is an essential feature of quantum theory" (*ibid.*, p. 380). The reconstruction starts therefore with two axioms: relativistic causality and non-locality – the first axiom corresponding to 'no-signalling' and the second being analysed in the neutral definition of non-local correlations in the sense of Bell's theorem. The study proceeds then by investigating which theories give rise to non-local correlations, while preserving causality. "Thus, our result is completely independent of quantum mechanics or any particular model" (*ibid.,* p. 381). The first immediate result of such investigation is the deduction that the two axioms can be kept compatible only in a non completely deterministic QT. "Then a 'negative' aspect of quantum mechanics – indeterminacy and limits on measurements – appears as a consequence of a fundamental 'positive' aspect: the possibility of nonlocal action" (*ibid.,* p.380).

---

[3] For a different interesting illustration of the limits of the C*-algebra approach, see also (Myrvold 2000).

After such a set up, it becomes more and more clear how the study of quantum non-locality is a central aim of the paper as much as the attempt of an axiomatization of QT. The paper in fact continues with an investigation of what kind of correlations one can get, that are consistent with the two principles, put forward through the manipulation of the toy models defined by these two principles. Such toy models (which are now called PR boxes, after the authors) consist in an imaginary device composed by two black boxes, one of which is given to Alice, the other to Bob. Both the experimenters can ask the box questions and receive answers, as the device has an input and an output port for each one of them. For each input, Alice receives a uniformly distributed random output and the same is the case for Bob; these outputs are non-locally correlated. However, PR devices do not support superluminal signalling between Alice and Bob, as the probability of Alice obtaining a given result *a* is independent of the question asked by Bob.

By playing with the possibilities provided by these PR boxes, Popescu and Rohrlich show that quantum correlations are not the sole non-local correlations that are consistent with such a setup, but that there exist other, so-called 'post-quantum', correlations that can be more non-local than quantum correlations[4]. In other words, quantum non-locality is not the sole kind of non-locality allowed in a world where the two principles of non-locality and 'no signalling' hold.

This discovery has opened the door to a series of questions about non-locality: if post-quantum correlations don't violate the 'no signalling' principle, why doesn't our world instantiate them? Why is our world only *this much* non-local, when it seems that it could be more?[5] It is this kind of question that is therefore addressed by PR models' explanations.

---

[4] Here non-locality is measured as the amount of violation of Bell's inequalities.
[5] From the point of view of someone used to the traditional interpretative questions, this question might seem odd. The issue here, one could claim, is not why the world does not instantiate correlations more non-local than QT, but how non-local correlations can even be possible! In considering this switch, however, don't forget that the possibility of post-quantum correlations would have important implication for information processing tasks. It is, therefore, essential to the achievement of the aims of this program to find out whether these correlations happen in our world, and, if not, why.

Building on Popescu's and Rohrlich's result, later works (Brassard et al. 2006, Brunner and Skrzypczyk 2009) suggest that the explanation of why our world only instantiates quantum non-locality, and no more non-locality than that, lies in another information-processing principle. Let's see how.

The availability of stronger correlations makes communication tasks easier, i.e. such correlations allow one to solve communication problems with the use of less resources. For instance, let's say that Alice and Bob have the task of calculating a Boolean function *f(x, y)*, where *x* is known only by Alice, and *y* only by Bob. Obviously, in order to solve the problem they must exchange some information. However, such information could be reduced depending on the availability of correlated systems shared by the two. It is, for instance, known that, in the calculation of some functions, sharing a pair of entangled systems could reduce the amount of information to be sent from one wing of the experiment to the other (Buhrman et al. 1988); however, for the calculation of some other functions (e.g. the inner product between two numbers) Alice will have to send to Bob the same number of bits as her input, whether or not the experimenters share a quantum entangled pair.

Brassard et al. (2006) have therefore shown that correlations that are more non-local than quantum correlations would make any communication complexity trivial, i.e. with them, any Boolean function could be calculated by Alice and Bob with the exchange of only one bit of information. Finally, Brunner and Skrzypczyk (2009) have shown that the same happens with post-quantum correlations that are arbitrarily close to classical correlations (i.e. arbitrarily close to the limit imposed by Bell's inequalities). In short, Brunner and Skrzypczyk show that from such post-quantum correlations it is possible to distil[6] correlations that are strong enough to make communication complexity trivial. The hope of the developers of this program is to show that this result is generalizable to all post-quantum correlations.

This suggests a possible reason why we have only so much non-locality in the world, as "most computer scientists would consider a world in which communication complexity is trivial to be as surprising as a modern physicist would find the violation of causality." (Brassard et al. 2006, p.2) They therefore put forward the conjecture that the explanation of the existent limit in the non-locality of our world lies in a new information-theoretic axiom about the impossibility of trivial communication complexity.

---

[6] Distillation is a procedure that begins with a large number of weakly correlated systems and ends up with a smaller number of strongly correlated systems, 'distilling' in this way non-locality.

## 4. Preliminaries

At first sight it might be unclear in what sense the above discussed accounts are explanations. In his 'manifesto' for axiomatic reconstructions of QT, Alexei Grinbaum (2007) summarizes the kind of explanatory reasoning that is supported by such theories, in the following way:

"- Why is it so?,

- Because we derived it.".

*Contra* Grinbaum, I will argue that a mere mathematical derivation does not account for the main and novel epistemic contribution provided by ARQIT explanations. In this section I will provide a preliminary analysis of such a contribution, which will establish the basic features that an adequate account of explanation in ARQIT will have to account for.

Researchers working in the field of information-theoretic physics often insist that quantum information theory implies a new perspective in our understanding of non-local entanglement "as a new kind of non-classical resource that could be exploited, rather than an embarrassment to be explained away" (Bub 2015). However, what is typically considered the most urgent foundational problem of quantum non-locality consists in describing the processes underlying the occurrence of quantum correlations. As such, the explanation of quantum correlations is part and parcel of the question 'how could the world be such that it behaves the way QT predicts?'. Under the minimal phenomenological interpretation of Shannon (and von Neumann) information as a measure of the amount of correlation between variables, an information-theoretic description of quantum correlations remains completely uninformative about how such correlations come about (Timpson 2013). Such a description in fact leaves open even the question of what kind of non-locality is involved here: merely apparent non-locality, as in Everettian interpretations, or a strong kind of non-locality, as in Bohmian theories.

Within the interpretation tradition, the switch of perspective suggested in the above quotation might therefore seem quite unhelpful – merely a way to replace the quest for a genuine explanation of phenomena with different, instrumentalist and technologically motivated questions (e.g. 'how can we use entanglement for information transfer tasks?').

However, an exclusive focus on the way QIT deals with questions whose natural locus is the interpretation program might cause one to miss the real novel explanatory contribution of ARQIT, of which some philosophers have instead caught a glance. For instance, in his brilliant book on quantum information theory and the foundations of QT, Timpson observes that: "[y]es, we would understand more about quantum mechanics when the measurement problem (etc.) were resolved; but we would also understand more about quantum mechanics if we were to know where the theory lies within a sufficiently broad space of physical theories of interestingly different kinds." (2013, p. 186).

As this passage suggests, one of the pivotal epistemic contributions of ARQITs is to generate just such a comparative understanding – an adequate account of explanation of ARQIT must therefore implement this kind of contribution. Focusing on this kind of understanding allows to see how ARQIT does not only address instrumentalist concerns and how, on the contrary, the central concerns of ARQIT are crucial questions in the foundations of QT, "those [questions] that have lain close to the heart of anyone interested in the foundations of quantum mechanics, since its inception: How does the quantum world differ from the classical [and postquantum, I add] one?" (Timpson 2008, p.2).

Different answers have been given to this question. For Planck, the distinctive feature of QT is the discretization of the energy levels of oscillators, for Bohr, it is the discretization of angular momentum, for de Broglie it is the wave nature of matter, for Schrödinger it is entanglement, for Heisenberg, it is the non-commutativity of the algebra of observables, and so on to indeterminism, superposition etc. According to John Wheeler the quantum principle was like the 'Merlin principle' the magician who could change form when pursued (Largeault et al. 1980).

Axiomatic reconstructions of QT address the question 'how does the quantum differ from the classical?' with the analytical tool of the axiomatization of the theory, which carries with it significant advantages. On the one hand, in fact, the axiomatic method provides a more objective and intersubjective standard for the evaluation of proposals: the core of QT is represented by the axioms of the best axiomatization of QT – where axiomatizations are evaluated according to typical standards (simplicity, comprehensiveness, informativity). On the other hand, the fact that a reconstruction is based on phenomenological principles whose meaning is easily accessible to scientists' understanding allows abstracting from possibly controversial features of the theory and guarantees more intersubjectivity in the understanding of the meaning of each of its elements.

In addition, ARQIT possesses a powerful abstracting tool, i.e. the notion of information as quantified by Shannon and von Neumann. We have seen that measures of information quantify the amount of correlation between variables, while abstracting from the constitution of the systems realizing, and from the processes yielding, such correlations. A description of the structure of reality in terms of information-processing protocols, therefore, allows an ontologically neutral but quantitatively precise description of the world in terms of the correlations between systems. Under these premises, a characterization of QT in terms of information-theoretic principles leads credence to the claim that the distinctive core of QT with respect to other theories lays in the specific kind of correlations that quantum systems realize. These considerations lead to another conclusion about the features of ARQIT explanations, which emerge exactly from such a new approach to the inquiry on the quantum world.

As I have said before, in the program of the interpretation of QT the main issue about non-locality consists in the explanation of how the correlations come about. In such a context, the question 'why is the quantum world non-local?' does not receive much attention, while some philosophers even argue that one should not expect to find an explanation to this question (e.g. Egg and Esfeld 2014). Contrary to this conclusion, ARQITs can address questions like 'why is the quantum world non-local?', or 'why is our world only this much non-local?'.

Returning to the discussion of the explanation of non-locality, if describing the world in terms of information processing allows us to isolate the features of QT that characterize the quantum against the classical world, then investigating non-locality in terms of the possibilities that it opens for information transfer highlights the way in which non-locality is linked to the very distinctive feature of the quantum world. But if this is so, then the way ARQIT's explanation works reflects a very intuitive and straightforward pre-theoretic sense of explanation, that is that *one explains a feature or a behaviour P of s by showing how P depends on the 'essence' of s*. According to the CBH reconstruction, therefore, non-local entanglement is explained as depending on the two no-go principles 'no broadcasting' and 'no superluminal information transfer'. In the PR boxes reconstruction, instead, non-locality is itself part of the distinctive feature of QT and the specificity of quantum non-locality depends on the conjunction with two other principles: 'no signalling' and 'non-trivial communication complexity'.

In what follows, I will propose an alternative account of explanation that successfully mirrors the above-seen pre-theoretical sense of explanation and accounts for the way in which this kind of comparative understanding is achieved.

## 5. An account of explanation in ARQIT

It could be claimed that the mere derivation, or unification of quantum phenomena under a few information-theoretic principles and a coherent formal structure already counts as a Deductive-Nomological explanation, or as an explanation by unification (Kitcher 1981). For instance, Flores (1999) characterizes explanations in theories of principle[7] (and therefore in ARQITs) as providing

---

[7] The now well-known distinction between theories of principle and constructive theories was originally drawn by Einstein: "We can distinguish various kinds of theories in physics. Most of them are constructive. They attempt to build up a picture of the more complex phenomena out of the materials of a relativity simple formal scheme from which they start out. Along with this most important class of theories there exists a second, which I will call 'principle-theories.' These employ the analytic, not synthetic, method. The elements which form their basis and

explanations by unification. However, neither the Deductive-Nomological, nor the Unificationist approach capture the specific contribution of ARQIT in the understanding of the quantum world. Such accounts of explanation, in fact, neglect both the comparative kind of understanding provided by ARQIT, and the fact that non-locality is explained by showing how it depends on what I called, in a pre-theoretic anticipation, the 'essence' of QT. That the Unificationist account cannot exhaust the explanatory contribution of ARQIT was already argued for instance in (van Camp 2011). Here it is argued that the most important explanatory contribution to be expected from ARQITs is that they delineate the preconditions for a (constructive) explanation of quantum phenomena. As argued by van Camp, the Unificationist (and, I add, the Deductive-Nomological) model, cannot account for this explanatory work. Although I agree with the analysis put forward by van Camp, I think that this is just part of the story and that the explanatory contribution provided by ARQIT goes beyond showing "that quantum mechanics requires conceptual clarification at a foundational level, determine in what respect it is required, and that its principles are ones which are constitutive of a coherent theoretical and conceptual framework whereby meaningful explanation is made possible." (van Camp 2011, p.9). As I will argue, ARQIT does not merely provide a sense of understanding, nor a heuristic guide for the achievement of the genuine explanation (provided only by a constructive interpretation of QT), but it provides genuine and self-contained explanations of some aspects of quantum non-locality.

*5.1 The bare bones*

In developing the details of my account, I take inspiration from the account of explanatory proof in mathematics formulated by Mark Steiner (1978).[8] Although Steiner's account was developed as an account of explanations in pure mathematics, its central idea that "to explain the behavior of an entity, one deduces the behavior from the essence or nature of the entity" (Steiner 1978, p.143), captures what I believe is the main explanatory content of ARQIT.[9] In his account Steiner replaces the notion of essence with the notion of *characterising property,* defined as a "property unique to a given entity or structure within a family or domain of such entities or structures" (Steiner 1978, p.143). According to Steiner's account, an explanatory proof:

---

starting-point are not hypothetically constructed but empirically discovered ones, general characteristics of natural processes, principles that give rise to mathematically formulated criteria which the separate processes or the theoretical representations of them have to satisfy" (Einstein, 1919, p. 228).

[8] Steiner's account has received criticisms as a theory of explanation in mathematics (e.g. (Resnik and Kushner 1987) and (Hafner and Mancosu 2005)). Such objections question the validity of Steiner's account as a universal account of mathematical explanation. Here, I do not defend either the universality, or the applicability in mathematics of the account, and for this reason I will not discuss these criticisms.

[9] It could be said that the mathematical derivations I described are, in the end, explanations in the Hempelian sense of mere derivation from Laws of Nature. However, being formulated as an account of explanatory proofs in pure mathematics (where every proof, whether explicative or not, is a derivation!) Steiner's account has the virtue, which will turn out very useful for my purposes, of showing what differentiates these accounts as explanatory in contrast with any generic mathematical derivation.

> "makes reference to a characterizing property of an entity or structure mentioned in the theorem, such that from the proof it is evident that the result depends on the property. It must be evident, that is, that if I substitute in the proof a different object of the same domain, the theorem collapses; more, I should be able to see as I vary the object how the theorem changes in response. In effect, then, explanation is not simply a relation between a proof and a theorem; rather, a relation between an array of proofs and an array of theorems, where the proofs are obtained from one another by the 'deformation' prescribed above." (p.144)

It should be already intuitively clear how this definition also mirrors the explanations illustrated in the previous sections.

First of all, the notion of characterising property, defined over a given family or domain, traces the comparative kind of understanding provided by ARQIT explanations. The definition of characterising property therefore adequately describes the (conjunction of the) principles of ARQITs, whose function is to isolate QT against a family of theories. The conjunction of the three CBH principles, for instance, isolates QT against the family of all theories representable with a C*-algebra, which CBH take to neutrally represent all mainstream physical theories. The principles of fantasy QT aim at discovering a set of principles (characterising properties) isolating QT against the family of all (existing or imaginary) non-local theories that don't allow superluminal signalling. As with Steiner's characterising property, ARQITs' characterising principles are "a derivative notion, since a given entity [theory] can be part of a number of differing domains or families. In fact, an object [a theory] might be characterized variously if it belongs to distinct families. Even in a single domain, entities [theories] may be characterized multiply" (*ibid.*, 144).[10] As we will see later in this section, the derivative character of this notion makes it especially useful in accounting for the scale of depth of ARQIT explanation.

Secondly, an essential part of the explanation consists in the derivation of the explanandum, which, given the set of QT principles (or axioms), is shown to be a theorem of QT. For instance, the explanation of non-local entanglement provided by the CBH reconstruction consists partly in making explicit how the existence of entangled states follows from 'no superluminal information transfer' and 'no broadcasting'. In the case of non-local boxes, part of the explanation why no stronger non-locality than quantum non-locality is instantiated in our world is constituted by the derivation of a contradiction between the existence of non-quantum correlations and a third (envisaged) information-theoretic principle about the impossibility of a world where communication complexity is trivial.

---

[10] Notice however that the concept of axiomatizing principles for a theory and that of characterising property are logically distinct. A characterising property does not necessarily entail every feature of the characterized object. It is sufficient that it isolate said object from the chosen family of reference. Not every characterising property, therefore, can play the role of a complete axiomatization of a theory.

Finally, for the comprehensiveness of the explanation it is crucial to show how, by changing the characteristic property, the theorem/explanandum changes in response. As argued by Steiner, the explanations of quantum non-locality provided by ARQITs do not only consist in the relation of mathematical derivability between non-locality and the principles, but rather in "a relation between an array of proofs and an array of theorems, where the proofs are obtained from one another by the 'deformation' prescribed above" (*ibid.*, p.143). In addition to this, it has to be said that the deformations of the principles that count as explanatory are not only those that lead to the derivation of new theorems, but are also those that, due to the reaching of a contradiction, lead to blocked derivations because the modified principles are incompatible. To include also these derivations in our explanatory information, I complete Steiner's criterion with the addition that the explanation also includes those modifications of the axioms that reach a contradiction and block the derivation of a modified theorem.[11]

Notice that, once applied to the physics context, the array of derivations translates into a pattern of counterfactual dependence between explanans and explanandum. More concretely, under the account I am proposing, ARQIT's explanation provides a specific kind of 'what-if-things-had-been-different' knowledge, where the counterfactual claims are produced by changing the theory's principles and exploring the kind of features that follow from such a deformation. In CBH, for instance, if 'no broadcasting' is dropped, then one has a classical phase space theory, while if the 'no superluminal information transfer' is dropped, one has a theory where distinct and distant physical systems are not kinematically independent, i.e. a strongly non-local theory. In the case of PR non-local boxes, the investigation of the models displaying non-locality and no superluminal signalling shows that the theories constrained by these two principles display a range of non-local correlations that go beyond those allowed by QT. In particular, once one considers the link between measure of non-locality and communication complexity, it seems that post-quantum correlations (and, in turn, any 'fantasy QT') would make communication complexity trivial. As noticed before, what-if-things-had-been-different knowledge is provided also when the derivation of a theorem is blocked because of the reaching of a contradiction, which shows that such a specific modification of the characterising property is not possible. For instance, in the case of PR non-local boxes a first immediate result is that a toy model that displays non-locality and no signalling can't also display complete determinism, otherwise the two axioms would be incompatible (Popescu and Rohrlich 1994, p. 380).

I have claimed before that in the account here offered, ARQIT does not merely increase our sense of understanding of QT, nor does it only play a heuristic function to guide the search for a genuine explanation, but it provides genuine explanations. It is therefore necessary to write a few words to justify this claim. First of all, many accounts of explanation currently tend to converge in the claim that the core of a scientific explanation lies in the counterfactual dependence between explanans and explanandum. Morrison (1999) for instance has argued that explanatory models exhibit a certain kind of structural dependence and, elaborating on this idea, Bokulich (2011) has argued that this structural dependence should be articulated in terms of counterfactual dependence. The same point has been made more recently by Reutlinger (2012) and Pincock (2014). The claim that information about counterfactual dependence is at the core of scientific explanation encompasses different kinds of scientific explanations, among them also causal or mechanistic explanation (Felline 2015). This is for instance clear in those theories of mechanisms that attribute a central role to counterfactual dependence in the definition of a mechanism (Craver 2007, Glennan 2010). The same kind of counterfactual knowledge is provided by ARQIT explanation. In practice, an ARQIT shows that if we lived in a world characterized by different quantum principles, then non-locality would change in such and such a way. The explanatory role of models is therefore linked to the fact that they provide information about a pattern of counterfactual dependencies or, in other words, they provide 'what-if-things-had-been-different' kind of information.

---

[11] I am thankful to an anonymous referee for pointing out the necessity of this clarification.

Within this framework, the difference between ARQIT's and other kinds of explanations is to be found in the different 'origin' of such a counterfactual dependence (Bokulich 2011). For instance, in a mechanistic explanation the explanandum counterfactually depends on a mechanical property of the explanans, *via* a relation of causal production. To make an example, the speed of my car is explained as depending on my pushing of the accelerator pedal, via the mechanical details of the motor inside the car, its intermediate components, their behavior and interactions. In an ARQIT, non-locality is explained as depending on the information-theoretic principles axiomatizing QT, via a relation of mathematical dependence between the formal representative of non-locality and the formal representatives of the information-theoretic principles.

Finally, although I am not going to push this last point, the formulation of the relation of explanatory relevance as a relation between the formal representatives of the explanans and the explanandum suggests that the most natural conceptual framework where to locate such an account of explanation is the epistemic view of scientific explanation. I take the latter as the view according to which a scientific explanation is a finite class of operations (inferences and other 'movements' of thoughts) on representations whose aim is to enrich our knowledge of an explanandum (Wright, manuscript)

A comprehensive understanding of the specific features of this kind of explanation requires a thorough investigation of the notion of characterising property. In the rest of this section I will underline some interesting features of this notion and neutralize two potential objections to the account of explanation here proposed.

*5.2 Two potential objections neutralized*

We have seen that ARQIT explains by deriving the explanandum from a characterising property and making explicit the counterfactual dependence between such a characterising property and the explanandum. We have also seen, however, that the same theory can have different characterising properties depending on the family over which the characterising property is defined (it is also possible that the same theory has more than one characterising property to isolate it against the same family). As a consequence, the same explanandum might have different potential explanations based on different characterising properties: depending on the family against which QT is characterized we could obtain different explanations of, say, entanglement. This raises the issue of multiplying explanations of the same explanandum and therefore the problem of evaluating which one is the best explanation. I argue, however, that this does not represent a pernicious problem for the proposed account of explanation, as it provides all the necessary tools for defining a sufficiently objective scale of depth for the evaluation of different explanations.

The rule that a world with fewer brute facts is more understandable than a world with more applies here as in virtually any variety of scientific explanation. In this sense, unification is a virtue of ARQIT's explanation, rather than its core: other things being equal, an explanation starting with fewer assumptions (and therefore a simpler characterising property) is to be considered better than one with more. The first feature that influences the depth of an explanation is therefore the simplicity of the characterising property.

In order to appreciate a second (probably more interesting) element that influences the explanatory power of ARQIT's explanation, consider the way such explanations work, namely by providing a kind of counterfactual knowledge. It follows from this that the kind of counterfactual information provided by the explanation affects the explanatory power of this explanation. In turn, the kind of counterfactual information provided by an ARQIT depends on the framework against which the theory is characterized (i.e. the family over which the characterising property is defined), as that family constrains the space of possibilities that are investigated with the modification of the characterising property. For instance, the generality of the family and its neutrality are features that influence the significance of the explanation: on the one hand, the more general the domain defining the space of possibilities, the more counterfactual information you can get; on the other hand, if the family of reference is more neutral, the space of possibilities explored is less 'informed' by the selection of the family, and more physically significant relations of dependence can be captured by the deformation of the characterising properties. As an illustration of how the neutrality of the family of reference influences the depth of an explanation, think about the case of the CBH reconstruction. We have seen how the choice of C*-algebra might be too strong, as it rules out by fiat an interesting class of theories (deterministic hidden variable theories). Furthermore, we have seen how, as a consequence of such an a priori selection of the family of reference, the claim that in the CBH characterization theorem the information-theoretic principles characterize QT is significantly reduced in scope. In this sense, an explanation showing how to deduce non-locality from the two axioms in the family of theories representable by a C*-algebra has less physical significance, and is less informative about the origin of non-locality, than an explanation using a more neutral framework.

This paper about explanation in ARQIT can't exhaust all the features that might contribute to the explanatory depth of an ARQIT – a thorough analysis of this point would risk overshadowing other elements of this proposal. However, the elements here outlined should be sufficient to support the claim that the potential multiplicity of explanations in ARQIT is not problematic, as the account outlines a clear and intersubjective way of evaluating and confronting explanations. Finally, notice that the same elements that contribute to a better explanation (simplicity, comprehensiveness, informativity) are mirrored by the elements that contribute to the physical significance of axiomatizations. The search for the best axiomatization of QT, therefore, goes hand in hand with the search for the best explanation.

There is another interesting consequence of the illustrated multiplication of explanations that deserves some attention. Consider that, in the same way as ARQITs single out QT with information-theoretic principles, we can also envisage the possibility of an axiomatization that reverses this order. For example, we can imagine an axiomatic system where, contrarily to the CBH reconstruction, the existence of entangled states is part of the axioms and 'no broadcasting' or 'no superluminal information transfer' are among the theorems. In general, while in the illustrated case studies a feature of QT is explained by appeal to information-theoretic principles, at the same time the proposed account allows as potential explanation also a derivation that reverses the explanatory order.

This consequence might seem disturbing – a replica of the problem of asymmetry originally formulated in the context of the DN and successively of the Unificationist model. However, as with the multiplication of explanations just analysed, this feature does not represent a problem for the account of explanation I am advocating. Take the well-known example of the flagpole and its shadow. In this example, the problem of asymmetry consists in the fact that, according to the DN (and the Unificationist) account, the length of a flagpole could explain the length of its shadow, as well as the other way around. On the other hand, while the height of a flagpole can clearly be used to explain the length of its shadow, the opposite explanation seems highly counterintuitive. An accredited diagnosis of the problem is that the analysis of this case misses the fact that the most natural explanation to ask for—of both the length of the shadow, and the height of the flagpole—is an explanation of *how* the shadow (or the flagpole) *is produced* in that specific length, and is therefore a causal explanation. Since causality is an asymmetrical relation, it therefore follows that if *a* causally explains *b*, then *b* cannot causally explain *a*. This asymmetry, though, is not a feature of the generic explanation, but originates with the (implicit) quest for a causal explanation – causality being therefore the origin of the asymmetry.

It might be objected at this point, that this limited example does not prove that there is no objective generic explanatory direction from the length of the flagpole to the length of the shadow. It might be that such an a priori explanatory direction exists and corresponds to the causal direction. This conjecture, however, has been already proven wrong. Take, for instance, the analysis of this same counterexample by Bas Van Fraassen, where the question 'why the flagpole is so long?' requires a functional explanation. In this case, the explanation is that the flagpole was built with the aim of casting a shadow of that exact length. In this version of the example, the length of the shadow is the explanans and the length of the flagpole is the explanandum, contradicting the claim that there is an objective explanatory arrow independent of the kind of explanation requested.

It can be said, therefore, that there are no reasons to assume that when *a* explains *b*, *b* cannot a priori explain *a*, but that such an asymmetry might instead be determined by the specific kind of explanatory relation that informs the explanation.

This being said, is asymmetry also a feature of the kind of explanation of non-locality provided by ARQIT? As we already know at this point, the answer to this question is 'no'. I have said that explanations provided by ARQITs (of 'why non-local entanglement?' or 'why this much non-locality?') explain by showing how the explanandum is a consequence of the characterising property of QT. We have also seen that the characterising property is a derivative notion. A theory can be characterized against a number of domains or families of theories. Even in a single family, theories may be characterized multiply. In particular, it might be that there are reconstructions where the existence of non-local entanglement is part of the axioms, and 'no broadcasting' is part of the theorems. Contrarily to the notion of cause or of function, there is no feature in the notion of characterising property that implies asymmetry, in the sense of a priori forbidding cases where the explanandum of one explanation become part of the explanans, and the explanans the explanandum.

Notice however, that the fact that our account does not need to discriminate a priori between *a* explaining *b* and *b* explaining *a* does not imply that such two potential explanations must be both accepted. It only means that the question of what is the *correct* explanation is a matter of investigation, which depends on what the best axiomatization is, with the criteria that we have already seen.

**6. ARQIT, ontology, and the interpretation of QT**

So far I have provided an account of what kind of explanatory gain we can expect from ARQITs. The main point of the previous section is that ARQITs can aspire at providing genuine explanations of some aspects of quantum non-locality. The aim of this section is to illustrate how such a claim can contribute to debates in the foundations of QT.

The significance of ARQIT in the foundations of physics has been articulated in a variety of different claims, each of which require a detailed and specific examination. In this section I will take into account and analyse specifically two claims.

A first assumption, perhaps vague but nonetheless intuitively strong, is that ARQIT's explanatory power must witness the special role that the structure of information, as constrained by the principles, must have in the ontology of QT, or that such explanatory power witnesses that QT is about quantum information, or even that, if QT is a fundamental theory, then "every item of the physical world has at bottom—at very deep bottom, in most instances— an immaterial source" (Wheeler 1990, p.35). The plausibility of this claim is probably reinforced by the unificatory power of the explanations provided by ARQIT. At a first sight, the intuitive idea of subsuming the structure of QT under few independent principles might seem the most concrete realization of the quest for the content of QT.

This conclusion, though, is unnecessary in order to account for the explanatory power of ARQIT and therefore, if we limit ourselves to considerations related to scientific explanation in ARQIT, unwarranted.

In order to understand why this is so, think again about the role of the principles in ARQITs' explanations of non-locality. First of all, they have to isolate QT with respect to the family of theories represented by C*-algebras – this function can be played by the principles without any foundational meaning. Secondly, they must logically (mathematically) imply the explanandum – they can do so, independently of the role of information in the foundations of QT. Finally, the explanation must show how the explanandum changes when the principles change – also in such a kind of counterfactual reasoning the ontological status of information plays no role.

In conclusion, this explanation can work even under a minimal phenomenological/instrumentalist interpretation of information. If we accept the provided account of the explanatory power of ARQIT, therefore, such an explanatory power is not to be cashed out in terms of a privileged ontological role of information-theoretic principles (aka characterising property) with respect to the derived structure of QT.

In order to see how this conclusion is essentially dependent on the account of explanation assumed, it is interesting to see how different accounts of explanations might lead to an opposite conclusion about the ontological significance of information theoretic principles.

One way of accounting for the explanatory power of principles and for their role in the explanation of other Laws of Nature, developed notably in (Lange 2011, 2013), is to ground such a role in metaphysical assumptions. According to Lange's metaphysics of Laws, natural necessity is hierarchically structured. A higher place in such a metaphysical hierarchy allows some Laws of Nature to act as constraints over other Laws of Nature. A constraint can explain the structure of lower-level Laws (and therefore phenomena that are instantiations of such a structure) because they show that such a structure is not accidental, but there is some sort of necessity in it. For instance, conservation laws are more necessary than dynamical laws, and therefore the former can explain the latter's structure. As a result of this strategy, Lange's account of the explanatory power of principles heavily depends on his metaphysical account of Laws and Necessity. If applied to ARQIT, such an account would imply that the explanatory role of information-theoretic principles with respect to the structure of QT would be conditional on their modal, metaphysical priority. To be sure, this would not necessarily imply claims such as that the physical world is made of information – however, it would surely warrant a heavily ontologically charged interpretation of Shannon information as a fundamental constituent of the physical world.

The evaluation of Lange's metaphysical account is outside the scope of this paper. Here, it suffices to say that the analysis I have proposed of the explanatory power of principles does not rely on specific metaphysical assumptions but rather focuses on the role of constraints as delineating an epistemic necessity. This, it has to be said, does not imply any straightforward conflict with Lange's metaphysical background – however it does not even require such a background.

Some philosophers take a stronger stance about the significance of ARQIT and maintain that, as a consequence of ARQIT, the search for a traditional interpretation of QT should be abandoned. Grinbaum (2007), for instance, describes a neat contrast between the two programs and argues that the shift to the reconstruction program should correspond to an abandonment of the interpretational program. The arguments he displays to support such a conclusion make appeal to broad considerations about the aims of physics and its relationship with philosophy. Although I find Grinbaum's arguments objectionable, I will not discuss them, as such broad arguments are outside the scope of this paper. In this section I will only evaluate the conjecture of the disjunctive relation between ARQIT and interpretation in terms of its consequences in the domain of explanation.

CBH (2003) have put this idea in especially clear and useful terms, by saying that ARQIT makes an interpretation of QT in terms of the behaviour of particles and waves 'explanatorily irrelevant'. In this section I test the acceptability of such a claim in the limited domain of the explanation of non-locality. I will therefore ask the question: can the ARQIT explanation of non-locality replace or screen-off an explanation of non-locality as provided by a traditional interpretation of QT?[12]

If we consider what has been said so far about ARQITs' explanations, then the answer to the above question should be negative: interpretations of QT address the explanation of non-locality as the question of *how* quantum correlations occur – the problem lying in the difficulty of providing a description of how the world is to be made in order to yield the correlations. The explanation of how a phenomenon occurs is usually taken to require an account in 'mechanical', 'dynamical' or more generically 'causal' (in the sense of productive causation) terms. The large majority of philosophers of physics argue that correlations should be explained in such a way, although the constantly growing material produced since the origins of QT in the attempt to formulate a causal account of quantum correlations has not yet achieved much consensus. Clearly, such an explanation cannot be given in terms of information.

However, causal explanation is not the only explanation-how. A possible alternative is to reject the assumption that quantum correlations should be explained causally and acknowledge that they should be explained 'structurally'. A structural explanation is an explanation where the explanandum is shown to be an instantiation of a fundamental structure of the world, rather than produced by some underlying processes (Dorato and Felline 2011, Felline 2011). Basically, this explanation shows that the explanandum is 'what happens when nothing happens'. The most well known example of a structural explanation is the explanation of length contraction in the special theory of relativity, where the former is explained by showing how it is not the result of a process undergone by material bodies, but rather an instantiation of the geometric structure of space–time, the latter being not Euclidean as previously expected, but Minkowskian.

The parallel with explanations in special relativity and in ARQIT was already proposed in CBH (2003) paper and has since then been central in Bub's work. Even Bub's most recent work exemplifies that something similar to a structural explanation is at the centre of his conjecture about explanation in ARQIT: "the conceptual revolution in the transition from classical to quantum physics should be understood as resting on the recognition that there is an information-theoretic structure to the mosaic of events, and this structure is not what Shannon thought it was" (Bub 2015,

---

[12] Considerations of this kind are extraneous to the literature devoted to PR boxes. This latter case study, therefore, will be left out of the discussion of this chapter.

p. 6).

If we follow such a conjecture, non-local quantum correlations, therefore, might be explained structurally as an instantiation of the structure of information that is not Shannon-like, but von Neumann-like.

In order for the structure of information to provide a structural explanation, however, it is necessary to go beyond the minimal phenomenological interpretation of information, and show that such a structure is fundamental in the sense that, as the geometry of spacetime, it is not explainable with, or inferable from, the dynamical or constitutive details of underlying particles or waves. Under the assumption that the structure of information is a fundamental structure (in the sense just explained) ARQIT can explain quantum phenomena as being the instantiation of a fundamental structure of the world. Also, it follows that no interpretation of QT in terms of particles or waves can provide another deeper explanation than this one exploiting the fundamental structure of information. On the contrary, without the assumption that the structure of information is a fundamental structure, there is always the possibility of a more fundamental story that infers/explains the structure of correlations, and we are back to square one. If such story is possible, then it has to provide an answer to the question: how quantum correlations come about?

Philosophers have taken different approaches to show that the structure of information indeed possesses such a fundamental status.

A first notable strategy is represented by the philosophical approach to quantum information theory called 'information immaterialism' (Zeilinger 1999), according to which quantum information is "the basic category from which all else flows" (Timpson 2010, p.209). The advocates of this view interpret the claim that the quantum state is about quantum information as the claim that information (the immaterial) is the fundamental subject matter of physical theory. Under the assumption that the world is, at its bottom, information, the information-theoretic description provided by ARQIT ceases to represent a mere tool for the quantification of the amount of correlation between unknown systems (and result of unknown processes) and represents instead the most fundamental description of the reality underlying said correlations. According to this view, other explanations of quantum phenomena in terms of particles or waves, collapse or non-collapse of the wave-function, are therefore ruled out as competitors of this account.

The ontological picture implied by information immaterialism still needs to be clearly spelled out and this makes a meaningful assessment of this strategy especially difficult. Some criticisms, in particular, seem to be hard to overcome. Among others, Timpson (2010, 2013) argues that this view rests on a conflation between the epistemic sense of information and the physical sense of information proper of Shannon information.

Like Zeilinger, also CBH argue that the conceptual problems of QT dissolve once one is ready to take the right philosophical lesson from quantum information theory: interpreting the quantum state as quantum information and taking the structure of information as fundamental. However, their argument (put forward in (Clifton et al. 2003) and elaborated more in detail in the succeeding work of Bub (2000, 2004, 2005)) is an epistemological argument rather than an ontological one. The CBH argument has influenced the debate over the philosophical meaning of ARQIT and more in general of quantum information theory in a way that goes beyond the significance of the CBH theorem itself. The details of such philosophical arguments deserve therefore a careful attention.

The way CBH argue for this conclusion is to conjecture that the CBH theorem makes any constructive mechanical interpretation of quantum theory in principle empirically underdetermined. In Bub's own words:

> "You can, if you like, tell a mechanical story about quantum phenomena (via Bohm's theory, for example) but such a story, if constrained by the information-theoretic principles, can have no excess empirical content over quantum mechanics, and the additional non-quantum structural elements will be explanatorily superfluous" (Bub, 2005, p. 14).

As a consequence, the rational epistemological stance is to reject any such interpretation as unacceptable, interpret QT as a theory of principle about the possibilities and impossibilities of information transfer and regard quantum information as a physical primitive (Bub, 2004). Bub devoted much of his subsequent work to the attempt to deepen and strengthen the claim that the CBH theorem made any constructive interpretation of QT empirically underdetermined; however, many philosophers criticize this claim. Besides the already mentioned (Timpson 2013) here it is worth mentioning that Duwell (2007) provides a careful survey and analysis of different arguments made by Bub to the conclusion that the CBH theorem makes information a physical primitive and rejects them. In another notable paper Hagar and Hemmo (2006) argue that quantum information theory is an essentially incomplete theory of quantum phenomena, requiring a further account in terms of a constructive QT with a clear solution of the measurement problem. Notice that their objection applies also to Zeilinger's immaterialism and in general to any 'black box' account of quantum phenomena. An interpretation of QT with a determinate stance with respect to the measurement problem is instead necessary in order to provide coherent predictions on the results of the experiment.

**Conclusions**

To sum up, the story here outlined has a positive and a negative side.

On the positive side, ARQIT can provide a genuine explanatory contribution to our understanding of quantum phenomena. Such an explanatory contribution does not merely consist in its heuristic role in the achievement of the (genuinely explanatory) constructive quantum mechanical theory, nor does it merely consist in answering technologically motivated questions. On the contrary, ARQIT's explanations aim at providing direct answers to questions that are central in the foundations of physics.

On the negative side, I have argued that such explanations do not exhaust the number of questions that arise from the investigation of quantum phenomena. Against the enthusiastic advocates of the 'paradigm shift' (Grinbaum 2007), the analysis proposed of the kinds of explanations provided by ARQITs still leaves space for an explanatory contribution by the interpretative program. In CBH terminology: ARQIT does not make a mechanical quantum theory explanatorily superfluous.

One of the main virtues of the philosophy of scientific explanation when applied to significant cases studies is its potential to clarify important features of such case studies and therefore to contribute to open issues in the foundations of physics. In this paper, I have provided a small sample of how such a contribution might work – it would therefore be incautious to draw general conclusions about QIT and the interpretation of QT. However, it is still useful to outline the direction where the analysis so far is pointing to.

The lesson that I take from what has been said here is that the explanatory role of information is much different than often implicitly assumed with scientific explanations, where the explanatory role of an entity (structure, property, law...) is typically taken to reflect a significant role in the ontology. On the contrary, the explanatory role of information consists partly in providing an ontologically neutral but quantitatively precise framework for the description of quantum correlations. But the explanatory role of information as a structure of probabilistic correlations also consists in picking up the specificity of QT against other classes of theories. The fact that QT can be characterized with few information-theoretic principles supports therefore the claim that the distinctive core of QT with respect to classical mechanics lies within the different structure of probabilistic correlations. On the other hand, the fact that the explanation of some quantum phenomena still requires an interpretation in terms of particles and waves, providing a solution of the measurement problem seems to witness against the claim that QT is about quantum information.

**Acknowledgements**

A special thanks to Giovanni Valente for his encouragement and suggestions for the improvement of this paper. The final version of this paper benefited from precious comments by two anonymous referees, but also from discussions with Jossi Berkovitz, Michael Cuffaro, Richard Healey and all the participants to the Information Theoretic Interpretations of Quantum Mechanics Workshop on June 11-12, 2016, at the Rotman Institute of Philosophy, in London, Ontario. Finally, I am extremely thankful to Angelo Cei, Lucas Dunlap, Mauro Dorato, Matteo Morganti and Giuseppe Sergioli for comments on earlier drafts of this paper.**References**

Bennett C.H. and G. Brassard. (1984). "Quantum cryptography: Public key distribution and coin tossing," in *Proceedings of IEEE International Conference on Comput- ers, Systems, and Signal Processing*, 175–179. IEEE.

Bokulich, A. (2011). "How scientific models can explain". *Synthese*, 180(1), 33–45.

Brassard, G. et al. (2006). "Limit on Nonlocality in Any World in Which Communication Complexity Is Not Trivial". *Physical Review Letters,* 96, 250401.

Brunner, N. and Skrzypczyk, P. (2009). "Nonlocality Distillation and Postquantum Theories with Trivial Communication Complexity" *Physical Review Letters,* 102, 160403.

Bub, J. (2000), 'Quantum Mechanics As A Principle Theory'. *Studies in the History and Philosophy of Modern Physics*, 31B, 75–94.

Bub, J. (2004). 'Why the Quantum?' *Studies in the History and Philosophy of Modern Physics*, 35B, 241–266. arXiv:quant-ph/0402149 v1.

Bub, J. (2005). 'Quantum Theory Is About Quantum Information', *Foundations of Physics*, 35(4), 541–560. arXiv:quant-ph/0408020 v2.

Bub, J. (2015). 'Quantum Entanglement and Information', *The Stanford Encyclopedia of Philosophy* (Summer 2015 Edition), Edward N. Zalta (ed.), URL = <http://plato.stanford.edu/archives/sum2015/entries/qt-entangle/>.

Bub, J. (2016). *Bananaworld: Quantum Mechanics for Primates*. Oxford University Press.

H. Buhrman, R. Cleve, and A. Wigderson. (1988). In *Proceedings of the 30th Annual ACM Symposium on Theory of Computing* (ACM, New York), p. 63.

Clifton, R., Bub, J., and Halvorson, H. (2003). "Characterizing quantum theory in terms of information theoretic constraints". *Found. Phys.*, 33(11):1561. Page refs. to arXiv:quant-26

ph/0211089.

Craver, C. F. (2007). *Explaining the brain*. Oxford University Press.

Dorato, M. and Felline, L. (2010). "Scientific Explanation and Scientific Structuralism" (in print), in A. Bokulich & P. Bokulich (eds.), *Scientific Structuralism*, Boston Studies in the Philosophy of Science, Springer.

Duwell, A. (2007). "Reconceiving quantum theory in terms of information theoretic constraints". *Studies in the History and Philosophy of Science Part B: Studies in the History and Philosophy of Modern Physics*, 38 (1), pp.181-201.

Egg, M., & Esfeld, M. (2014). "Non-local common cause explanations for EPR". *European Journal for Philosophy of Science*, *4*(2), 181-196.

Einstein, A. (1919). "Time, space, and gravitation". *The London Times*, November 28. Page references to reprint (under the title: What is the theory of relativity?) in A. Einstein, *Ideas and Opinions* (pp. 227–232). New York: Crown Publishers, 1954.

Felline, L. (2011). Scientific Explanation between Principle and Constructive Theories. *Philosophy of Science*, 78(5), 989-1000.

Felline, L. (2015). Mechanisms meet structural explanation. *Synthese*, 1-16. doi:10.1007/s11229-015-0746-9

Flores, F. (1999). 'Einstein's theory of theories and types of theoretical explanation', *International Studies in the Philosophy of Science*, 13:2, 123-134.

Glennan, S. (2010). 'Mechanisms, causes, and the layered model of the world', *Philosophy and Phenomenological Research*, 81(2), 362-381.

Grinbaum, A. (2007) "Reconstruction of Quantum Theory" *British Journal for the Philosophy of Science,* 58(3), 387-408.

Hafner, J. and P. Mancosu, (2005). "The Varieties of Mathematical Explanation". In P. Mancosu *et al*. (eds.), *Visualization, Explanation and Reasoning Styles in Mathematics*, Berlin: Springer, 215–250.

Hagar, A., and Hemmo, M. (2006). "Explaining the unobserved—Why quantum mechanics ain't only about information". *Foundations of Physics*, 36(9), 1295-1324.Page refs. to http://arxiv.org/abs/quant-ph/0512095.

Kitcher, P. (1981). Explanatory unification. *Philosophy of science*, 48(4), 507-531.

Largeault, J. et al. (1980). *A question of Physics: Conversations in Physics and Biology*. University of Toronto Press.

Morrison, M. (1999). Models as autonomous agents. In M. Morgan & M. Morrison (Eds.), *Models as mediators: Perspectives on natural and social science* (pp. 38–65). Cambridge: Cambridge University Press.

Mermin, N. D. (2004). "Could Feynman Have Said This?", *Physics Today*, 57, 10 – 2.

Myrvold, W. C. (2010). "From physics to information theory and back." In A. Bokulich, G. Jaeger (eds.) *Philosophy of quantum information and entanglement*, 181-207.

Pincock, C. (2014). "Abstract explanations in science". *The British Journal for the Philosophy of Science*. doi: 10.1093/bjps/axu016.

Popescu, S. (2006) "Quantum mechanics: Why isn't nature more non-local?" *Nature Physics* 2, 507 – 508.

Popescu, S. and Rohrlich, D. (1994). "Quantum nonlocality as an axiom". *Foundations of Physics*, 24(3): 379-385.

Resnik, M., and D. Kushner. (1987). "Explanation, Independence, and Realism in Mathematics". *British Journal for the Philosophy of Science*, 38: 141–158.

Reutlinger, A. (2012). "Getting rid of interventions". *Studies in History and Philosophy of Science Part C: Studies in History and Philosophy of Biological and Biomedical Sciences*, *43*(4), 787-795.

Rovelli, C. (1996). "Relational quantum mechanics". *International Journal of Theoretical Physics*, 35(8):1637–1678.

Steiner, M. (1978). "Mathematical Explanation*". Philosophical Studies*, 34:2 p.135.

Timpson, C. (2008). "Philosophical aspects of Quantum Information Theory" D. Rickles (ed.) The Ashgate Companion to the New Philosophy of Physics (Ashgate). Page refs. To: arXiv:quant-ph/0611187v1.

Timpson, C. G. (2010). "Information, immaterialism, instrumentalism: Old and new in quantum information." In A. Bokulich, G. Jaeger (eds.) *Philosophy of quantum information and entanglement*, 208-227.

Timpson, C. G. (2013). *Quantum information theory and the foundations of quantum mechanics*. Oxford University Press.

Van Camp, W. (2011). "Principle theories, constructive theories, and explanation in modern physics". Studies in History and Philosophy of Science Part B: Studies in History and Philosophy of Modern Physics, 42(1), 23-31.

Wheeler, J. A. (1990). "Information, physics, quantum: The search for links". In Zurek, W., editor, *Complexity, Entropy and the Physics of Information*, 3–28. Addison- Wesley, Redwood City, CA.

Woodward, J. (2003). "Scientific Explanation". *The Stanford Encyclopaedia of Philosophy*. Edward N. Zalta (ed.). URL = http://plato.stanford.edu/entries/scientific-explanation/

Wright, C. D. (manuscript) "Scientific Explanation: Mechanistic, Epistemic, Ontic".

Zeilinger, A. (1999). "A foundational principle for quantum mechanics". *Found. Phys.*, 29(4): 631–43.